\documentstyle[preprint,aps,epsfig]{revtex}

\preprint{\begin{tabular}{r}
MPI-PhT/2001-14 \\
hep-ph/0105288
\end{tabular}}

\tightenlines

\relax
\newcommand{\TeV}{\,{\rm TeV}}

\newcommand{\MeV}{\,{\rm MeV}}

\newcommand{\eV}{\,{\rm eV}}

\newcommand{\beqa}{\begin{eqnarray}}
\newcommand{\eeqa}{\end{eqnarray}}
\newcommand{\ba}{\begin{array}}
\newcommand{\ea}{\end{array}}
\def\321{$SU(3)\times SU(2)\times U(1)$}

\def\a{\alpha}
\def\b{\beta}

\def\d{\delta}

\def\e{\epsilon}
\def\l{\lambda}

\def\m{\mu}

\def\zi{\xi}
\def\r{\rho}
\def\s{\sigma}
\def\t{\tau}
\def\th{\theta}

\def\beq{\begin{equation}}
\def\eeq{\end{equation}}
\def\ol{\overline}

\def\bmat{\left( \begin{array}}
\def\emat{\end{array} \right)}
\def\barr{\begin{eqnarray}}
\def\earr{\end{eqnarray}}

\def\l{\left}
\def\r{\right}

\def\lsim{\raise0.3ex\hbox{$\;<$\kern-0.75em\raise-1.1ex\hbox{$\sim\;$}}}
\def\gsim{\raise0.3ex\hbox{$\;>$\kern-0.75em\raise-1.1ex\hbox{$\sim\;$}}}

\def\dmsq{\Delta m^2}
\def\s33{_{3 \times 3}}
\def\s3k{_{3 \times k}}
\def\sk3{_{k \times 3}}

\def\s44{_{4 \times 4}}
\def\s4k{_{4 \times k}}
\def\sk4{_{k \times 4}}

\def\chie{\chi_e}
\def\chimu{\chi_\mu}
\def\chitau{\chi_\tau}
\def\chiperp{\chi_\perp}
\def\chipos{\chi_+}
\def\chineg{\chi_-}
\def\eps{\epsilon}

\begin{document}

\title{Neutrino anomalies and large extra 
dimensions}

\author
{Amol S. Dighe$^1$\footnote{amol.dighe@cern.ch} and 
Anjan S. Joshipura$^2$\footnote{anjan.joshipura@cern.ch} \\[.5cm]}
\address
{$^1$Max-Planck-Institut f\"ur Physik, F\"ohringer Ring 6, 
D-80805 M\"unchen, Germany. \\
$^2$Theoretical Physics Group, Physical Research Laboratory, \\
Navarangpura, Ahmedabad, 380 009, India.}
\maketitle

\bigskip

\begin{abstract}

Theories with large extra dimensions can  generate 
small neutrino masses when the standard model neutrinos are
coupled to singlet fermions propagating in higher dimensions.
The couplings can also generate mass splittings and mixings
among the flavour neutrinos in the brane.
We systematically study the minimal scenario involving 
only one singlet bulk fermion coupling weakly to the
flavour neutrinos.
We explore the neutrino mass structures in the brane that
can potentially account for the atmospheric, solar and
LSND anomalies simultaneously in a natural way.
We demonstrate that in the absence of a priori mixings
among the SM neutrinos, it is not possible to reconcile
all these anomalies. 
The presence of some structure in the mass matrix of the
SM neutrinos can solve this problem.
This is exemplified by the Zee model, which when embedded 
in extra dimensions in a minimal way can account for
all the neutrino anomalies.

\end{abstract}

\newpage

\section{Introduction}

The present data from the experiments on atmospheric, solar and 
reactor neutrinos indicate neutrino flavour oscillations. 
The data from each of these sets of experiments individually
can be explained by a single dominating mass square difference
$\dmsq$ and a mixing angle $\theta$.
The atmospheric neutrino data \cite{atm}
indicate $\nu_\mu \leftrightarrow \nu_\tau$ as the dominant mode,
with 
$\Delta m^2_{atm} = (1 - 8) \cdot 10^{-3} \mbox{ eV}^2 ~,~
\sin^2 2\theta = 0.8 - 1.0$.
There is no compelling evidence that the electron neutrinos
participate in the oscillations of atmospheric neutrinos. Moreover,
the CHOOZ experiment \cite{chooz} gives an upper 
bound on the mixing of $\nu_e$: we have
$\sin^2 2 \theta_e \leq 0.1$
where $\nu_3$ is the mass eigenstate such that 
$\dmsq_{31} \approx \dmsq_{32} \approx \dmsq_{atm}$.
The three MSW solutions (LMA, SMA and LOW) as well as the vacuum 
oscillations can provide reasonable fits to the solar neutrino data 
\cite{solar}, all these solutions have
$\dmsq_\odot \leq 2 \cdot 10^{-4}$ eV$^2$. 
The results of the LSND experiment \cite{lsnd} are neither
confirmed nor fully excluded by the KARMEN2 data \cite{karmen}, 
and the combined fit allows a region \cite{lsnd-kar}
$\Delta m^2_{LSND} = (0.1 - 1) \mbox{ eV}^2 ~,~
\sin^2 2\theta_{\mu e} \approx 10^{-3} - 10^{-2}$.

In the context of only three known neutrino species, 
the $\dmsq$s corresponding to the solutions of the three
neutrino anomalies (atmospheric, solar and LSND) cannot
be reconciled and if the oscillation mechanism is to be used to 
explain all the anomalies\footnote{Some non-oscillation 
``exotic'' solutions have been proposed \cite{exotic}, 
but we shall not discuss them here.},
the introduction of a sterile neutrino becomes
necessary. There are two simple schemes of mixing among four neutrinos
which can account for all the neutrino anomalies:  

(i)
In the ``3+1'' scheme \cite{31}, the masses of the three active states are 
separated from that of the sterile one by the LSND scale. 
The possible $\nu_e-\nu_\m$ 
oscillations seen at LSND arise through simultaneous mixings of $\nu_e$
and $\nu_\m$ with the sterile neutrino $\nu_s$. 
Such a picture is strongly constrained
by laboratory disappearance experiments. It was argued \cite{31} 
that these experiments constrain the  $\nu_e-\nu_\m$
mixing below the experimental signal at LSND. This has changed
with the new LSND data and the 3+1 scheme is claimed \cite{no-31} to
be  viable for some parameter range, though
the detailed statistical analysis in \cite{grim}
does not support this claim. 

(ii)
The second scheme --- known as the ``2+2'' scheme \cite{31,22} --- 
corresponds to two
nearly degenerate pairs of neutrinos separated by the LSND mass scale. 
The mixings within the pairs account for the solar and atmospheric 
flux deficits. The small
oscillation probability seen in the LSND experiment can be explained
by requiring that $\nu_e$ and $\nu_\m$ belong to two different
degenerate pairs. This implies that the sterile state has significant
admixture either with $\nu_e$ or $\nu_\mu$ and thus plays an important 
role in generating either the solar or the atmospheric
neutrino deficit. The data on atmospheric
neutrinos disfavour $\nu_\m-\nu_s$ as 
the dominant source of the $\nu_\m$ 
oscillations \cite{atm}. 
The global fit to the available sets of the solar
neutrino data are also better described in terms of 
$\nu_e$ conversion to active rather than the sterile state, 
although the latter possibility 
is not completely ruled out \cite{ggarcia}. 
The 2+2 scheme will be ruled out if both the solar and
the atmospheric neutrinos are found to be mainly oscillating
to active neutrinos.

It follows that both the  above schemes are strongly constrained and may
not prove adequate for incorporating the 
LSND signal. In this case, the description of data 
in terms of four neutrino mixing would be challenged and more complicated
possibilities would be needed. 
Quite apart from phenomenological incompatibility, the very
existence of a light sterile state needs theoretical justification. 
While there have been attempts in this direction \cite{qgf},
the introduction of more than one light sterile states 
can add to theoretical difficulties.

The above theoretical and phenomenological problems find natural
solutions in theories
with extra dimensions of mm size \cite{extrad}. 
Firstly, such theories can provide justification for the existence
of  light sterile states which form a part of the  Kaluza Klein
(KK) tower associated with a singlet fermion residing in the bulk
volume of the full higher dimensional theory.
In addition, the presence of more than one light sterile state
allows additional phenomenological possibilities not envisaged
in the context of the  2+2 or 3+1 schemes mentioned above. 
As a result, these theories provide a  potentially
useful framework for understanding all neutrino anomalies.

The standard model particles are assumed to be confined
to a three-brane embedded in a higher dimensional bulk. The gauge
singlet states are not confined this way and propagate in the bulk.
Such theories have been argued to provide
some understanding of neutrino masses if one assumes the presence of
a singlet neutrino $\nu_B$ propagating in the bulk. The couplings
of $\nu_B$ to the flavour states residing in bulk are suppressed by a
volume factor corresponding to the extra dimensions and lead to very
small neutrino masses \cite{small-numass,others}
in spite of the new physics at TeV scale. 

The minimal approach consists of assuming only five 
dimensions\footnote{ The observed accuracy of Newton's law
over earth-sun distance requires the presence of two or
more extra dimensions.
We have implicitly assumed that the radii of all other extra dimensions
are much smaller than that of the fifth dimension. 
One can work in this case with an effective five dimensional
theory.} 
and the presence of only one bulk
neutrino coupling  to all the known flavors through  
\beq
\label{ma-def}
{m_\a\over v} \bar{l_\a} \phi \nu_B+h.c. ~,
\eeq
where
\beq
\label{ma-estim}
m_\a\equiv{h_\a v M_f\over M_P}\approx 6\cdot 10^{-5} \eV h_\a \left ({
M_f\over \TeV}\right) \quad .
\eeq 
Here $h_\alpha$ is the Yukawa coupling, and $M_f (M_P)$ is
the fundamental (Planck) mass scale.
Neutrino masses and mixings are determined
by four parameters: three mixing parameters 
$\zi_\a\equiv \sqrt{2}m_\a R$
and an overall mass scale $m^2\equiv m_e^2+m_\m^2+m_\tau^2$.

From the point of view of the 4-dimensional theory, the $\nu_B$ residing
in a five dimensional world contains a massless state which can provide
a sterile neutrino needed to understand various neutrino anomalies. 
The $n^{\rm th}$ KK
excitation associated with $\nu_B$ has a mass $m_n = n/R$. 
For $R\sim 0.1~mm$, masses of some of the low lying excitations 
would be in the range needed for the 
occurrence of the MSW resonance for solar neutrinos. 
The mixing of the KK states
with $\nu_e$ is governed by $\zi_e$. For $M_f$ in the TeV
range, this mixing is in the  range needed for the small mixing angle
solution of the solar neutrino problem. Hence, the MSW solution in this
context is correlated to observable extra dimensions and 
TeV mass scale \cite{ds}.
However, these values of the parameters  cannot
provide the atmospheric neutrino scale if Yukawa couplings $h_\a<1$. 
This makes it difficult to reconcile
all the neutrino anomalies and one needs to go beyond this
minimal picture. The following different possibilities have been 
considered in the literature:

({\em i}) Extension of the minimal scheme with three bulk
neutrinos was considered in \cite{mo-ba}.
The phenomenology of this case is described in terms of an arbitrary
mixing
matrix, three Dirac brane-bulk masses $m_\a$ and corresponding
$\nu_\a-\nu_{B_\a}$ mixing parameters $\xi_\a$. The solar and atmospheric
neutrino deficit can be explained for small $\xi_\a$. But this requires 
larger $m_\m$ than the natural value expected from (\ref{ma-estim}), 
and consequently the fundamental scale of about three
to four orders of magnitude higher than TeV. 
On the other hand, for larger $\xi_2$
and/or $\xi_3$, the atmospheric neutrinos oscillate to the unfavoured mode
of the sterile state. Moreover, one cannot explain the LSND
results in a satisfactory manner in 
any of these cases \cite{mo-ba}.

({\em ii}) 
Lukas {\it et al.} \cite{lukas1} considered an extension which
allows for a mass term of the bulk neutrino. Such an extension
is compatible with all neutrino anomalies provided
one introduces three bulk neutrinos and lepton number violation
in the brane-bulk couplings. Some of these couplings are required
to be as large as $m_\alpha \sim 10^{-1}$ eV as in the case ({\em i}). 
Thus either one needs a relatively large fundamental scale or 
Yukawa couplings much larger than 1. 

({\em iii}) The neutrino masses and mixings exclusively
occur due to physics in the bulk in both the above cases. It is 
conceivable that physics in the brane itself may have some seed for
the neutrino oscillations to occur. This possibility was considered
in \cite{brane-mix}. It was shown that if the brane neutrinos have some masses
then their coupling to the bulk neutrino can induce mixings,
and hence oscillations, among the brane neutrinos.
In this context, models have been considered \cite{ling}
that try to solve some of the neutrino anomalies.

The aim of this paper is to study the feasibility of the suggestion
({\em iii}) from the point of view of solving all neutrino anomalies.
Motivated by the observation that strong mixing of sterile state with
$\nu_\m$ is disfavoured \cite{atm}, we shall work in the 
limit of only small
brane-bulk mixing, i.e. small $\zi_\a$. Neutrino oscillations
can be largely described in this framework from an effective $4\times 4$
matrix in the space of three active neutrinos and the massless mode
of $\nu_B$. One can use the standard seesaw approximation to determine 
this matrix provided one allows non-leading terms in this approximation.
We present the general formalism for this in Sec.~\ref{formalism}. 
Non-zero brane mass
for the electron neutrino can significantly influence the solution to
the solar neutrino problem proposed in \cite{ds}. We discuss this effect
quantitatively in Sec.~\ref{nuemass}. These results are then used  
in Sec.~\ref{unmixed} to study phenomenology of the minimal case in which 
one allows neutrino
masses but no mixing in the brane. It is shown that one cannot solve
all the neutrino anomalies in this case. In Sec.~\ref{zee}, we
study a specific structure of the neutrino mass matrix
based on the Zee model \cite{zee} and show that 
this, when embedded in the extra dimensions,
can explain all the neutrino anomalies. 
The last section gives summary of our results.

\section{General formalism}
\label{formalism}

\subsection{The Lagrangian}
\label{lagrangian}

We start by taking only one singlet neutrino in the bulk, with
the free bulk action given by
\beq
S_{\rm bulk} = \int d^4 x ~dy~ \ol{\Psi} \gamma^A ~i\partial_A \Psi~~,
\label{sbulk}
\eeq
where $\gamma^A ~~ (A = 0,1,2,3,4)$ are the 5-dimensional Gamma
matrices.
Here $x \equiv (x^0,x^1, x^2,x^3)$ are the usual 4-d coordinates,
and the SM brane is at $y \equiv x^4 = 0$.  
Let the bulk neutrino couple to 
all the known active flavours,
so that the brane-bulk Yukawa coupling is\footnote{
Since $\Psi$ and $\overline{\Psi}$ have the same transformation
properties under 5 dimensional Lorentz symmetry, we have 
chosen to name as $\Psi$ the bulk spinors with lepton number +1
(as in \cite{lukas2}), so that the brane-bulk coupling 
(\ref{sbrane}) is lepton number conserving.
Lepton number violating terms of the form 
$h_\alpha^c \ol{L}_\alpha H \Psi^c$ are possible in principle,
but we can get rid of them through the imposition of
a ${\bf Z}_2$ symmetry \cite{lukas2}. The same symmetry also
forbids a possible Dirac mass term $\mu \overline{\Psi} \Psi$ in
(\ref{sbulk}).}
\beq
S_{\rm brane} =  - \int_{y=0} d^4 x ~
\l( \frac{h_\alpha}{\sqrt{2 \pi R}} \frac{M_f}{M_P}~ 
\ol{L}_\alpha H \Psi + h.c. \r) ~~.
\label{sbrane}
\eeq 
The four dimensional effective action may be obtained by 
expanding $\Psi$ as
\beq
\Psi(x,y) = \frac{1}{\sqrt{2 \pi R}} \sum_{n \in {\bf Z}}
\Psi_n(x) exp \l( \frac{i n y}{R} \r)~~,
\label{psi-expand}
\eeq
where each of the modes $\Psi_n$ may be written in terms of
two 4-d left-handed Weyl spinors $\xi, \eta$ as
\beq
\Psi_n = \l( \begin{array}{c} 
i\xi_n \\ \bar{\eta}_n
\end{array} \r)~~.
\label{weyl}
\eeq
The effective 4-d Lagrangian involving KK modes can be written as
\beq
-{\cal L}_{mass} =
\sum_\alpha m_\alpha \nu_\alpha (\sum_{n \in {\bf Z}} \eta_n) +
\sum_{n \in {\bf Z}} m_n \xi_n \eta_n + h.c. 
\label{lmass-exp}
\eeq
where $m_n \equiv n/R$.
Defining new fields 
\beq
\xi^{\pm}_n  \equiv  (\xi_n \pm \xi_{-n})/\sqrt{2} ~~,~~
\eta^{\pm}_n \equiv  (\eta_n \pm \eta_{-n})/\sqrt{2} ~~,
\label{xietapm}
\eeq
we can write the above Lagrangian (\ref{lmass-exp}) as
\beq
-{\cal L}_{mass} =  
\sum_\alpha m_\alpha \nu_\alpha 
(\eta_0 + \sqrt{2} \sum_{n \in {\bf N}} \eta^+_n) +
\sum_{n \in {\bf N}} m_n (\xi^-_n \eta^+_n + \xi^+_n \eta^-_n)
+ h.c. 
\label{lm-final}
\eeq
The fields $\xi^+_n$ and $\eta^-_n$ decouple completely 
from the rest.

We add to the above Lagrangian a general mass term for the flavour
neutrinos
$\nu_\a ~~(\a=e,\m,\t)$\footnote{
This term explicitly violates the lepton number in the brane,
but it is still allowed under the ${\bf Z}_2$ symmetry that forbids
the bulk-brane lepton violating coupling \cite{lukas2}.}: 
\beq
\label{branemass}
{1\over 2} \sum_{\alpha,\beta}  \mu_{\alpha\beta} \nu_\alpha \nu_\beta
~~.
\eeq
The net Lagrangian  relevant for the neutrino oscillations can then be
written as
\beq
-{\cal L}_{mass} =  
\frac{1}{2} (\nu_e, \nu_\mu, \nu_\tau, \eta_0, 
\hat{\eta}, \hat{\xi})^T
~{\cal M}_\nu~ 
(\nu_e, \nu_\mu, \nu_\tau, \eta_0, \hat{\eta}, \hat{\xi})
\label{mnu-def}
\eeq
where
\beq
\hat{\eta} \equiv (\eta^+_1, \eta^+_2, ...) \quad,\quad
\hat{\xi}  \equiv (\xi^-_1, \xi^-_2, ...) 
\label{xieta-hat}
\eeq
The Majorana mass matrix ${\cal M}_\nu$ is
\beq
{\cal M}_\nu =
\bmat{cccccc}
\mu_{ee} & \mu_{e\mu} & \mu_{e\tau} & 
m_e & \sqrt{2} M_e & 0\\
\mu_{e \mu} & \mu_{\mu\mu} & \mu_{\mu\tau} & 
m_\mu & \sqrt{2} M_\mu & 0\\
\mu_{e \tau} & \mu_{\mu\tau} & \mu_{\tau\tau} & 
m_\tau & \sqrt{2} M_\tau & 0\\
m_e & m_\mu & m_\tau &  0 & 0 & 0 \\
\sqrt{2} M_e^T & \sqrt{2} M_\mu^T &
\sqrt{2} M_\tau^T & 0 & 0 & M_{\eta\xi} \\
0& 0 & 0 & 0 & M_{\eta\xi} & 0
\emat ~~,
\label{mnu}
\eeq
where the matrices $M_\alpha$ and $M_{\eta\xi}$ 
are defined as
\beq
M_\alpha \equiv m_\alpha (1,1,1,...) \quad , \quad
M_{\eta\xi} \equiv \frac{1}{R}{\rm Diag}(1,2,3,...) ~.
\label{sub-mnu}
\eeq

\subsection{Diagonalization of the neutrino mass matrix}
\label{diagon}

We now use the standard seesaw approximation to diagonalize 
eq.(\ref{mnu}). This may be written in the block form
\beq
{\cal M}_\nu = \bmat{cc}
m_L  & m \\
m^T & M 
\emat~,
\label{mnu-box}
\eeq
where 
\beq
m_L \equiv \bmat{cccc}
\mu_{ee} & \mu_{e\mu} & \mu_{e\tau} & m_e \\
\mu_{\mu e} & \mu_{\mu\mu} & \mu_{\mu\tau} & m_\mu\\
\mu_{\tau e} & \mu_{\tau\mu} & \mu_{\tau\tau} & m_\tau \\
m_e & m_\mu & m_\tau &  0 
\emat \quad,\quad
m \equiv \bmat{cc}
\sqrt{2} M_e & 0 \\
\sqrt{2} M_\mu & 0 \\
\sqrt{2} M_\tau & 0 \\
0 & 0
\emat \quad,\quad
M \equiv \bmat{cc}
0 & M_{\eta\xi} \\
M_{\eta\xi} & 0 
\emat~~.
\label{sub-box}
\eeq
The matrix $M$ is diagonalized by the unitary matrix $\hat{Y}$:
\beq
\hat{Y} = \frac{1}{\sqrt{2}} \bmat{cc}
1 & -1 \\ 1 & 1 \emat 
\quad {\rm so~ that } \quad
\hat{M} \equiv \hat{Y}^T M \hat{Y} = 
\bmat{cc} 
M_{\eta\xi} & 0 \\ 0 & -M_{\eta\xi} \emat ~~.
\label{y-hat}
\eeq
Then the action of the matrix
$ Y \equiv \bmat{cc} 1 & 0 \\ 0 & \hat{Y} \emat$
on ${\cal M}_\nu$ will be
\beq
\hat{\cal M}_\nu \equiv Y^T {\cal M}_\nu Y = \bmat{cc}
m_L & \hat{m} \\ \hat{m}^T & \hat{M} \emat
\quad {\rm where} \quad
\hat{m} \equiv m \hat{Y} = \bmat{cc}
M_e & -M_e \\ M_\mu & -M_\mu \\ M_\tau & -M_\tau \emat
\label{mnu-hat}
\eeq
The matrix $\hat{\cal M}_\nu$ may be ``block-diagonalized''
in general by
\beq
W \equiv \bmat{cc}
\sqrt{1 - B B^\dagger} & B \\
-B^\dagger & \sqrt{1 - B^\dagger B} \emat,
\quad {\rm  such ~that} \quad
\tilde{\cal M}_\nu \equiv W^T \hat{M}_\nu W =
\bmat{cc}
\tilde{m}_L & 0 \\ 0 & \tilde{M} \emat.
\label{W-def}
\eeq
Here, in the seesaw approximation 
that is valid for $\mu_{\alpha\beta}, m_\alpha \ll 1/R$,
we may expand the relevant quantities as \cite{grimus}
\barr
\tilde{m}_L & = & m_L - \hat{m} \hat{M}^{-1} \hat{m}^T 
- \frac{1}{2}(\hat{m} \hat{M}^{-2} \hat{m}^T m_L+ h.c.) + ... 
\label{mltilde} \\
\tilde{M} & = & \hat{M} + 
\frac{1}{2}(\hat{m}^T \hat{m} \hat{M}^{-1} + h.c.) + ... 
\label{Mtilde} \\ 
B & = & \hat{m} \hat{M}^{-1} + m_L \hat{m} \hat{M}^{-2} + ... 
\label{Bdef}
\earr
In the above expressions,
we have retained the non-leading terms neglected in
the usual seesaw approximation. This is necessary because of the fact that
the leading term
\beq
\hat{m} \hat{M}^{-1} \hat{m}^T = 0
\label{vanish}
\eeq
vanishes identically as can be checked from eqs. (\ref{sub-box}) 
and (\ref{mnu-hat}). The  non-trivial correction to $m_L$ is thus
provided by the third term in eq.(\ref{mltilde}) and the $\tilde{m}_L$  
is given by
\barr
\tilde{m}_L  & = & m_L
- \frac{1}{2}(\hat{m} \hat{M}^{-2} \hat{m}^T m_L+ h.c.)
\nonumber \\
 & = & m_L
- \frac{1}{2} \l[ \bmat{cccc}
\chi_e^2 & \chi_e \chi_\mu & \chi_e \chi_\tau & 0 \\
\chi_e \chi_\mu & \chi_\mu^2 & \chi_\mu \chi_\tau & 0 \\
\chi_e \chi_\tau & \chi_\mu \chi_\tau & \chi_\tau^2 & 0 \\
0 & 0 & 0 & 0 \emat 
m_L + h.c. \r]
\label{mlt-general}
\earr
where we have defined 
$\chi_\alpha \equiv \sqrt{\sum \frac{1}{k^2}} \xi_\alpha
=  m_\alpha R \sqrt{\sum \frac{2}{k^2}}$
for future convenience. 

Let $\hat{K}$ be the matrix that diagonalizes $\tilde{m}_L$
through $\hat{K}^T \tilde{m}_L \hat{K} = {\rm Diag} (m_1,m_2,m_3)$
and $V$ be the matrix that makes the elements of 
$\tilde{M}$ positive:
\beq
V \equiv \bmat{cc}
1 & 0 \\ 0 & i \emat
\quad {\rm so ~that} \quad
V^T \tilde{M} V = \bmat{cc}
M_{\eta\xi} & 0 \\ 0 & M_{\eta\xi} \emat~.
\label{V-def}
\eeq  
The matrix $\tilde{\cal M}_\nu$ is then completely
diagonalized by
\beq
K \equiv \bmat{cc}
\hat{K} & 0 \\ 0 & V \emat
\quad {\rm through} \quad
K^T \tilde{\cal M}_\nu K = {\cal M}_D~.
\label{K-def}
\eeq
The net mixing matrix which diagonalizes ${\cal M}_\nu$
is then
\beq
U = Y W K = \bmat{cc} 
\sqrt{1 - B B^\dagger}~ \hat{K} & B V \\
- \hat{Y} B^\dagger \hat{K} & 
\hat{Y} \sqrt{1 - B B^\dagger}~ V \emat~.
\label{U-def}
\eeq
To the first order in $m_\alpha R$, we have
\beq
B \approx \hat{m} \hat{M}^{-1} = R 
\bmat{c} m_e \\ m_\mu \\ m_\tau \\0 \emat
\bmat{cc} M_{1/n} & -M_{1/n} \emat
\quad {\rm where } \quad
M_{1/n} \equiv  \l( 1, \frac{1}{2}, \frac{1}{3},... \r)~.
\label{B-exp}
\eeq
Since $B \sim m_\alpha R \ll 1$, we may expand in powers of
$B B^\dagger$ to write the flavor states 
($\nu_e, \nu_\mu, \nu_\tau$ and $\nu_s \equiv \eta_0$) as
\barr
\nu_\alpha & = & \sum_\beta [\delta_{\alpha\beta} 
-\frac{1}{2}(B B^\dagger)_{\alpha\beta}] \hat{K}_{\beta i} \nu_i
+ \sum_{n \in {\bf N}} B_{\alpha n} (\eta^+_n + i \xi^-_n) 
\nonumber \\
 & = & \sum_\beta [\delta_{\alpha\beta} 
-\frac{1}{2}(B B^\dagger)_{\alpha\beta}] \hat{K}_{\beta i}\nu_i
+ \sqrt{2} \sum_{n \in {\bf N}} B_{\alpha n} N_n 
\label{nu-alpha}
\earr
where $N_n \equiv (\eta^+_n + i \xi^-_n)/\sqrt{2}$.
Note that
\beq
\sqrt{2} B_{\alpha n} = \left\{ \begin{array}{ll}  
\xi_\a / n \quad \quad & 
\alpha \in \{e,\mu,\tau \} \\
0 & \alpha = s \end{array} \right.
\label{B-alphan}
\eeq
so that the
mixing between the KK excitations and the $\nu_\a,\eta_0$
is of ${\cal O}({m\over M})$ as in the standard seesaw approximation,
and
\beq
(B B)^\dagger_{\alpha \beta} = \left\{ \begin{array}{ll}  
\chi_\alpha \chi_\beta \quad \quad & 
\alpha,\beta \in \{e,\mu,\tau \} \\
0 & \alpha = s~ {\rm or}~ \beta = s \end{array} \right. ~~.
\label{BBdag}
\eeq
To the first order in $m_\alpha R$, the approximation
\beq
U_{\alpha i} \approx \hat{K}_{\alpha i}
\label{u-k}
\eeq
is valid for $\alpha \in \{e,\mu,\tau,s\}$ and
$i \in \{ 1,2,3,4 \}$.

Eq. (\ref{nu-alpha}) coincides with the corresponding equation in case of
a single flavour derived for example in \cite{ds}. The mixing of
higher KK states with the brane states is unaffected by the
presence of masses in the brane. In contrast, the mixing among 
four neutrinos as determined by $\hat{K}_{\b i}$ in (\ref{nu-alpha}) 
is strongly dependent on the presence of neutrino masses in the brane.
In particular, the mixing among brane neutrinos may be generated
entirely from their couplings to the bulk neutrino. We shall study
implications of this in Sec~\ref{unmixed}.

\section{Electron neutrino mass and solar anomaly}
\label{nuemass}

The conversion of $\nu_e$ from the sun to sterile neutrinos 
can occur in this theory through the resonance of $\nu_e$ 
with the tower of the KK states with masses
${n\over R}$. This has been explored in detail \cite{ds} 
in the case of a massless electron neutrino.
With $R \sim 10^3$ eV$^{-1}$, the KK states are separated by
mass differences of the order of $10^{-3}$ eV, so
the $\nu_e$ mass $\lambda \lsim 10^{-3}$ eV will still not
affect the $\nu_e$ survival probability calculated therein.
However, in some of the scenarios that we will be considering  
in this paper,
$\nu_e$ (more precisely, the mass eigenstate with a considerable
fraction of the electron flavour) can have a non-zero mass 
$\lambda \geq 10^{-3}\eV$.
We discuss in this section modifications introduced
in the treatment of \cite{ds} due to such a mass. In particular,
relatively large $\nu_e$ mass can spoil the  solar neutrino 
solution proposed therein.

For an electron
neutrino with energy $E$, the density of the layer of
resonance with the $n^{\rm th}$ KK state is
\beq
\rho_n = \left( \frac{n^2}{R^2} - \lambda^2 \right)
\frac{m_N}{2 E G_F (Y_e - Y_n/2)}~,
\label{rho-n}
\eeq
where $m_N$ is the nucleon mass, $G_F$ is the Fermi coupling constant,
and $Y_{e(n)}$ is the number of electrons (neutrons) per nucleon
in the medium.
Eq.(\ref{rho-n}) gives the set of energies $E_{nR}$ beyond which 
$\nu_e$ undergoes resonance with the $n^{th}$ KK state:

\beq
E_{nR}\approx 0.3 \MeV \cdot  (n^2-\lambda^2 R^2) \left({10^6\eV^2\over
R^2}\right)~, \label{enr}\eeq 
The values of $E_{nR}$ can be appropriate for solar neutrino
provided $R\sim 10^{3} \eV^{-1}$. 
The specific excitations which participate in the
resonance are influenced by the value of $\lambda R$.  

The neutrino conversions 
take place mainly in the resonance layers (\ref{rho-n}). 
Since the mixing angle is small $(\xi_e/n \ll 1)$, the resonance 
layers are well separated. As the neutrinos travel outwards 
from their production point inside the sun, they encounter
the resonance layers corresponding to the densities (\ref{rho-n}).
The $\nu_e$ survival probability after passing through the
resonance layer is
\beq
P_n \approx {\rm Exp} \left[ -
\frac{\pi}{2} \left(\frac{n^2}{R^2} - \lambda^2 \right)
\frac{2 \xi_e^2}{ n^2 E} 
\left. \frac{\rho}{d\rho/dr} \right|_{\rho = \rho_n} 
\right] ~~.
\label{p-n}
\eeq
The net $\nu_e$ survival probability is
\beq
P_{surv} = \prod_{n=n_{min}}^{n_{max}} P_n
\label{psurv}
\eeq
where $n_{max}$ corresponds to the highest density resonance that
the neutrinos with energy $E$ encounter, and $n_{min}$ is 
determined by the condition $n \geq \lambda R$.
Since $P_n$ approaches 1
rapidly $(P_n \sim {\rm Exp} [-C/n^2])$ with increasing $n$,
the upper cutoff in $n$ is not of much practical significance.
The lower cutoff however may have a significant impact on the final
$\nu_e$ survival probability, depending on the $\nu_e$ mass.

For $\lambda=0$, the value of $P_n$ is independent of the specific
excitation $n$ which takes part in the resonance \cite{ds}.
This gets altered when $\lambda$ is of the order of the MSW scale or 
larger. In particular, a substantially large value for $\lambda$ 
implies large $n_{min}$ for the resonance and 
the suppression of $P_{surv}$ due to $P_n$s with low 
values of $n$ is
absent. This implies a higher $\nu_e$ survival probability.

\begin{figure}[htb]
\begin{center}
\epsfig{file=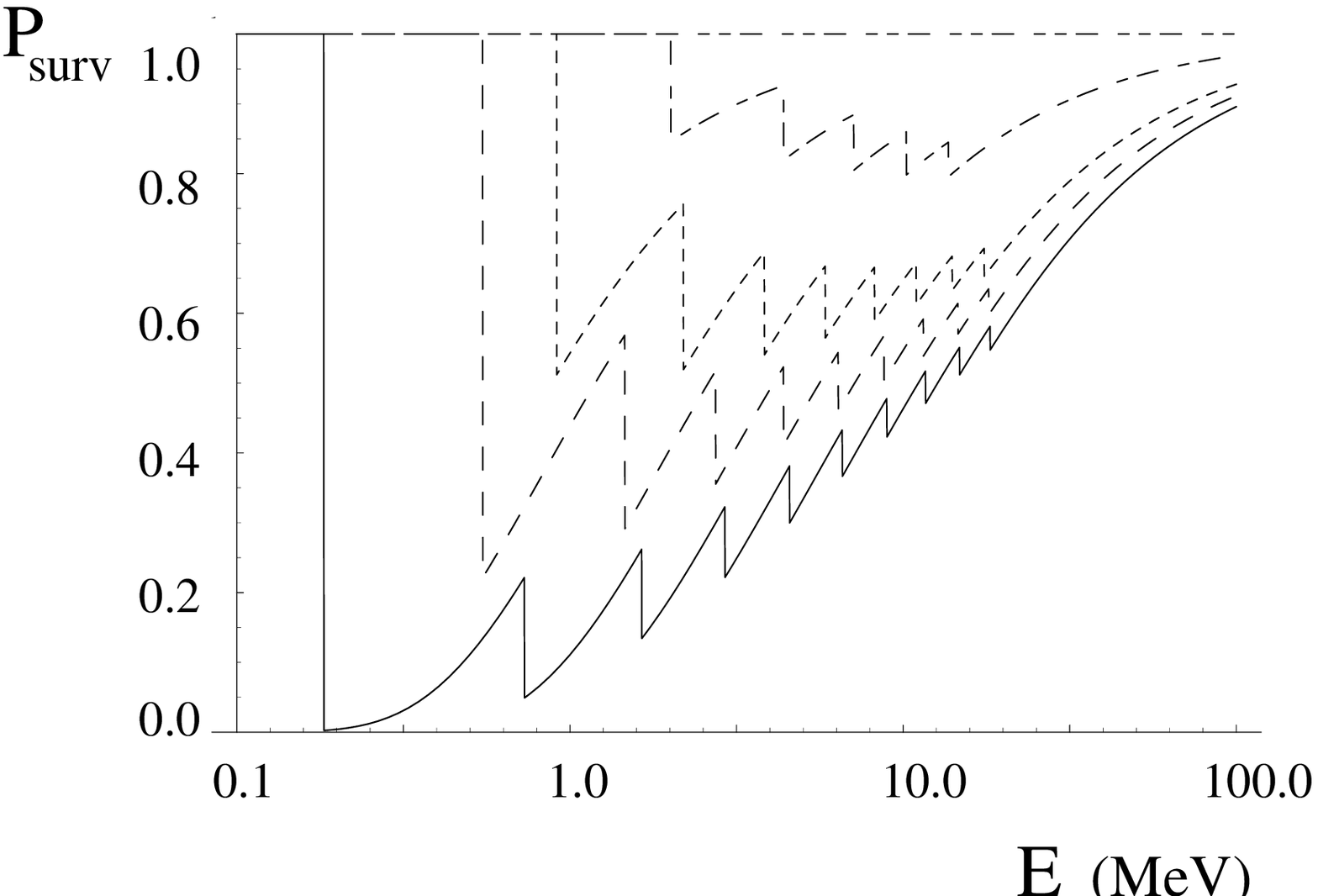,height=3in}
\end{center}
\caption{Survival probability of $\nu_e$ from the sun as a function
of energy. The (solid, long dashed, short dashed, long-short dashed,
long-short-short dashed) curves are for 
$\lambda = (0,1,2,5,10^3)\cdot 10^{-3} \eV$.
The purpose of this plot is only to illustrate the dependence of 
$P_{surv}$ on $\lambda$.
The details of neutrino production and transport inside the sun and the
energy resolution have not been taken into account. 
These will result in a smearing of the energy dependence shown in
the figure.
\label{surv-fig}}
\end{figure}

We show the value of $P_{surv}$ as a function of $\nu_e$ energy 
in Fig.~\ref{surv-fig} for different values of $\lambda$ using the parameters
$$4 \xi_e^2 = 10^{-3}~,~ R = 10^3 \eV^{-1} ~,~ Y_e = Y_n = 0.5 ~,~
\rho/(d\rho/dr) = 1.4 \cdot 10^{15} \eV^{-1}~~.
$$
The value of $P_{surv}$ is seen to be significantly affected by 
a non-zero $\lambda$.
In particular, it becomes $\sim 1$ for $\lambda\sim \eV$.
In this case, the resonance with the KK states
is ineffective in converting the solar neutrino with small mixing $\xi_e$.
We shall encounter different cases considered here in the subsequent
phenomenological analysis.

\section{Unmixed active sector}
\label{unmixed}

In this section, we discuss specific forms of $m_L$ which correspond
to massive but unmixed neutrinos in the brane.
Mixing among them can arise only indirectly through their couplings
to the bulk neutrinos. This possibility was proposed in \cite{brane-mix}
where the neutrino oscillation patterns  were studied numerically in this
scenario. We give below an analytic discussion and concentrate on
the feasibility of  solving all neutrino anomalies in this context.
Many of the important features needed for this can be elucidated
by considering only two generations. We thus first consider 
the case of two active flavours $\a$ and $\b$.

\subsection{Two generations with Majorana mass}
\label{2-gen}

We take the following form for the flavour mass matrix $\m_{\a\b}$:
\beq
\left(
\ba{cc}
\mu_1&0\\
0&\mu_2\\ \ea \right)~~.
 \eeq
The $3\times 3$   Majorana mass matrix
in the basis of the three
``flavor states'' $\nu_\alpha, \nu_\beta, \nu_s$ is given by
\beq
m_L = \bmat{ccc} \mu_1 & 0 & m_\alpha \\ 
0 & \mu_2 & m_\beta \\
m_\alpha & m_\beta & 0  \emat ~.
\label{ml-2}
\eeq
After taking into account the mixing with the bulk modes,
the mass matrix $\tilde{m}_L$ becomes (see eq.(\ref{mlt-general})) 
\beq
\tilde{m}_L = \bmat{ccc}
\mu_1 (1 - \chi_\alpha^2) & 
-(\mu_1 + \mu_2) \chi_\alpha \chi_\beta /2 & m_\alpha t \\
-(\mu_1 + \mu_2) \chi_\alpha \chi_\beta /2 &
\mu_2 (1 - \chi_\beta^2) & m_\beta t \\
m_\alpha t & m_\beta t & 0 \emat
\label{mlt-2}
\eeq
where $t \equiv 1 - \chi_\perp^2/2$ with 
$\chi_\perp \equiv \sqrt{\chi_\alpha^2 + \chi_\beta^2}$.

As expected, the bulk states have generated mixing among the 
flavour states. The corresponding mixing angle is seen to be given in the
limit $\mu_{1,2} \gg m_{\alpha,\beta}$ by
\beq
\tan 2 \theta_{\a\b}=-{(\mu_1+\mu_2) \chi_\a \chi_\b
\over (\mu_2-\mu_1)- (\mu_2\chi_\b^2-\mu_1\chi_\a^2)}~~.
\label{tan2theta}
\eeq
This induced mixing among flavour state can be large if 
$\mu_1 \approx \mu_2$
while it tends to be small for 
$\mu_1 \approx -\mu_2$. This is to be contrasted
with the pseudo-Dirac case where large mixing is linked to the presence
of almost equal and opposite masses. 
The mixing angle $\theta_{\a\b}$ is maximal in the
limit $\{ \mu_1= \mu_2, \chi_\a= \chi_\b \}$. 
The brane neutrinos are degenerate in this case but 
this degeneracy is lifted by couplings to the bulk neutrino. Thus
both the mass splitting and the (large) mixing can be completely
attributed to the higher dimensional physics in this case.
The mass squared difference among the three neutrinos can be worked 
out from (\ref{mlt-2}) and are given by
\beq
\label{deltas2}
\Delta m^2_{12}= 2 \m^2(- \chi_\perp^2 + \frac{1}{4} \tan^2 2 \phi)~~,~~
\Delta m^2_{31}\sim\Delta_{32}\sim \m^2~~,
\label{dmsq-2}
\eeq
where 
\beq
\m\equiv\mu_1= \mu_2~~,~~
\tan 2 \phi=-{2 r_\perp t \over 1- \chi_\perp^2}~~,~~
r_\perp \equiv \frac{\sqrt{m_\alpha^2 + m_\beta^2}}{\mu} ~.
\eeq

If the flavour states $\{ \a,\b \}$ are identified with 
$\{ \nu_\m,\nu_\tau \}$,
the the large mixing in (\ref{tan2theta}) 
can be identified with the mixing observed
in the atmospheric anomaly. The relevant (mass)$^2$ difference is given by
$\Delta m^2_{12}$ in (\ref{deltas2}).
A potentially interesting range of parameters is
$\m\sim 1 \eV$ and $\chi^2\sim 10^{-3}$. This range reproduces the
atmospheric
mass scale, has the required large mixing, and also has 
the seed of explaining LSND result through $\m\sim \eV$ when
the coupling to a nearly massless $\nu_e$ is turned on. 
The 3$\nu$ extension of this scheme is discussed in
Secs \ref{0mumu}.

Let us consider the alternative possibility corresponding to
$\{ \a=e,\b=\m \}$. Depending on the ranges of parameters, particularly
the mass $\m$, solution of the
solar neutrino problem in this context has
several interesting aspects:

\noindent ({\em i}) 
If $\mu$ is chosen to be around the MSW scale or lower 
($\mu^2 \lsim 10^{-6}$ eV$^2$), 
then $\nu_e$ can get converted to the KK excitations 
through MSW resonance with them, as discussed in Sec.~\ref{nuemass}.
Even if the conversion probability is somewhat reduced when
$\mu \sim$ MSW scale, such conversions can still contribute to
the SMA solution of the solar neutrino problem for small 
values of $\chi_e$. In addition,
another conversion mechanism becomes possible simultaneously if
$\chi_e\sim \chi_\m$ as would follow for non-hierarchical Yukawa couplings.
The $\nu_e - \nu_\m$ mixing angle is large in this case, and
$\Delta m^2_{12}$ in (\ref{dmsq-2}) can be appropriate for the vacuum solution
when $\chi_\perp^2\sim 10^{-4}$ (which also corresponds to the SMA
solution).
Thus the solar neutrino flux gets altered inside the Sun through the
resonance with KK states
and outside through vacuum oscillations with $\nu_\mu$. 
The simultaneous presence of
these two solutions helps in getting better agreement 
with the solar data since it 
is known that the vacuum oscillations of $\nu_e$ to sterile state
gives a poor fit to the rates observed in different solar neutrino experiments.
But this in conjunction with the SMA MSW resonance with
the KK excitations is argued \cite{msw+vac}
to describe various features of the solar neutrino data.
However, note that since both the mass scales generated here correspond to
$\dmsq_\odot$, it is not possible to incorporate the solutions to both
the atmospheric and LSND anomalies through the introduction of 
a single additional neutrino $\nu_\m$.

\noindent ({\em ii})
If $\mu$ is chosen around the LSND scale,
the KK resonance is ineffective as discussed in Sec.~\ref{nuemass}.
The solar neutrino problem can still find an explanation. This is because the 
$\Delta m^2_{12}$ controlling the solar oscillations can now be 
in the MSW range if $\mu^2$ is towards the
lower end of the LSND region and mixing parameter 
$\chi_\perp^2 \sim 10^{-3}$. 
Since the mixing angle $\theta_{e\mu}$ is large (\ref{tan2theta}),
this can provide the large angle MSW solution,
which is preferred by the solar data. The
higher dimensional physics is not directly involved in solar
neutrino problem in this case but its role is to generate mixing and 
mass splitting among
the active neutrinos. The oscillations relevant for 
the LSND solution could
occur indirectly through coupling with $\nu_s$. 
Incorporating the atmospheric neutrino problem in this context would 
require introducing the third neutrino $\nu_\tau$.
This extension is discussed in Sec.~\ref{mumumu}.

If one or more of the masses $\mu_1, \mu_2$ is zero, the mixing angle
generated (\ref{tan2theta}) is very small. Since we need at
least one large mixing angle (for solving the atmospheric
neutrino problem), we shall consider
only those scenarios in which at least two neutrinos have a 
nonzero degenerate mass to begin with. If we do not introduce any  scale
other than the common mass 
$\m$, we are led to two different possibilities corresponding to
(a) two degenerate and one massless neutrino (see Sec.~\ref{0mumu}), and
(b) all three degenerate neutrinos (see Sec.~\ref{mumumu}).
The complete neutrino mass spectrum in these models is 
determined in terms of five parameters: the common mass
$\m$ of neutrinos, three mixing parameters $\xi_\a$ and the 
compactification radius $R$. While their magnitudes are arbitrary,
we will concentrate on the consequences that follow when they are assumed
to be around the following "natural" values:
\beq
\label{values}
R^{-1}\approx (2-3)\cdot 10^{-3}\eV~~~;~~~
\xi_\a\equiv \sqrt{2} m_\a R\approx (0.03-0.4)~~~;~~~ 
\m^2\approx (0.1-1) \eV^2  ~. 
\eeq 
The value of $R$ is near the observational limit and it also
allows the possibility of an MSW resonance between $\nu_e$ and the tower
of the KK states \cite{ds}. Given this value and assuming 
the fundamental scale $M_f$ to lie
in $1-10 \TeV$ range, one obtains the quoted values for $\xi_\a$.
Value of $\xi_e$ in this range leads to a SMA solution to the solar neutrino 
problem. Natural value of $m_\a\sim 10^{-4}-10^{-5}\eV$ cannot help
in generating the LSND or atmospheric mass scale. 
The value of $\m$ therefore needs to be 
chosen near the LSND scale. Given this, the atmospheric scale follows 
naturally as we will see. The solar scale is explained in terms of the
value of $R$ chosen in the above equation.

\subsection{Three unmixed generations with Majorana masses
$\{0,\m,\m \}$}
\label{0mumu}

We consider the case corresponding to three unmixed neutrinos having the
masses $\{ 0,\mu,\mu \}$ in the brane. The 
$4\times 4$ mass matrix involving $\nu_e,\nu_\m,\nu_\tau,\nu_s$ is given
by
\beq
m_L =\bmat{cccc}
0 & 0 & 0 & m_e \\
0 & \mu & 0 & m_\mu \\
0 & 0 & \mu & m_\tau \\
m_e & m_\mu & m_\tau & 0 \emat \quad .
\label{ml-mumu}
\eeq
Starting with this $m_L$ and including the corrections due
to seesaw approximation as in eq.(\ref{mlt-2}), 
we get the following effective mass matrix:
\beq
\tilde{m}_L = \m\bmat{cccc}
0 &  - \chie \chimu/2  & -\chie \chitau/2 & \tau \chie \\
-\chie \chimu/2 &  1 - \chimu^2   &  -\chimu \chitau &  \tau \chimu \\
-\chie \chitau/2 & -\chimu \chitau &   1 - \chitau^2 &  \tau \chitau \\
\tau \chie &  \tau \chimu & \tau \chitau & 0  \emat~~.
\label{mlt-mumu}
\eeq
where  
$\tau \equiv (\sqrt{2  \sum (1/k^2)}~\m R)^{-1} (1 - \sum \chi_\a^2/2)$.

We can study consequences of the above equation approximately by
retaining quadratic terms in parameters $\chi_\a,\tau$. This approximation is
seen to be quite good for the choice of parameters as in eq.(\ref{values}).
The diagonalizing
matrix is given in this approximation by
\beq
\hat{K} = R_{23}(\theta)~ R_{34}(\phi)~ R_{13}(\psi) ~R_{14}(\omega)~,
\label{k-mumu}
\eeq
where $R_{ij}$s are $4 \times 4$ rotation matrices in the 
$i-j$ plane, and the angles are given by 
\beq
\theta = \tan^{-1}\frac{\chimu}{\chitau} ~,~
\phi = - \tau \chiperp ~,~ 
\psi = - \frac{\chie \chiperp}{2} ~,~
\omega =  \pi/4
\label{angles-mumu}
\eeq
where $\chiperp \equiv \sqrt{\chimu^2 + \chitau^2}~$.
The mass eigenvalues are 
\beq
\lambda_1 = - \mu \tau \chie ~,~
\lambda_2 = \mu ~,~
\lambda_3 = \mu( 1 - \chiperp^2) ~,~
\lambda_4 = \mu \tau \chie~.
\label{lambda-mumu}
\eeq
Note that two of the eigenvalues are degenerate.
They will be split by terms higher order in $\chi^2$.
By keeping the higher order terms in the diagonalization,
one can  show that the 1-4 splitting 
is of ${\cal O}(\m^2\delta^6)$ where $\delta$ denotes typical
magnitude of $\chi_\m\sim \chi_\tau\sim \chi_e \sim \tau$. We thus have
the following (mass)$^2$ differences:
\beqa
\dmsq_{24} &\approx& \dmsq_{34} \approx \dmsq_{21}
\approx \dmsq_{31} = \mu^2 ~,~ \nonumber \\
\dmsq_{23} &\approx& 2 \mu^2 \chiperp^2 ~,~\nonumber \\
\dmsq_{14}&\approx& {\cal O}(\delta^6\mu^2) ~.
\label{dmsq-mumu}
\eeqa

\begin{figure}[htb]
\begin{center}
\epsfig{file=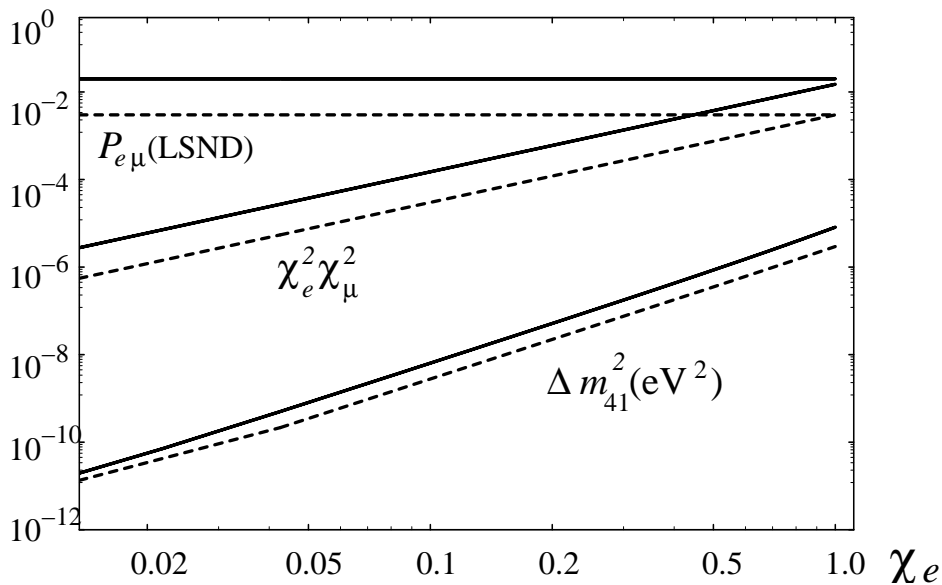,height=3in}
\end{center}
\caption{ { $\dmsq_{41}$ in $\eV^2$ (the lower pair of curves) and  
the predicted LSND probability (eq.~\ref{lsnd-mumu})
$\chi_e^2\chi_\m^2$ (the middle pair of curves) plotted against $\chi_e$.
The solid (dotted) curves correspond to $\mu^2=1.0 \; (0.2) \eV^2$.
The uppermost horizontal lines are experimental LSND probabilities .02
and .003 corresponding to  $\m^2=1.0$ (solid) and $0.2 \eV^2$ 
(dotted) respectively. } 
\label{lsn}}
\end{figure}

The $\dmsq_{23}$ which plays the role of $\Delta_{atm}^2$
describes the splitting between (almost) degenerate 
$\nu_\m-\nu_\t$ pair.
The $\nu_e-\nu_s$ pair is almost maximally mixed and 
is separated from the $\nu_\m-\nu_{\tau}$
pair by the LSND scale. 
This pattern thus reproduces the 2+2 scheme
of neutrino mixing. 
The phenomenology of the solar
neutrino however differs considerably from the 2+2 model.
Now the zero mode of $\nu_B$ as well its KK excitations
contribute to the solution of the
solar neutrino problem: the former through
non-zero $\dmsq_{14}$ and the latter ones through MSW resonances
with $\nu_e$. This is quantitatively displayed in Fig.~\ref{lsn}. 
This figure shows the variation of  $\dmsq_{14}$ 
obtained by diagonalizing
eq.(\ref{mlt-mumu}) with $\chi_e$ for different values of $\m$. 
The $\chi_{\m,\tau}$ are determined
by identifying $\dmsq_{23}$ and $\theta$ (see eq.~\ref{angles-mumu})
with the atmospheric
neutrino mass scale and mixing angle. It is seen that the favourable
value of $\chi_e\sim 0.02 $ for obtaining the SMA solution also
corresponds to $\dmsq_{14}\sim 10^{-10}-10^{-11}\eV^2$. 
Since the $\nu_e - \nu_s$ mixing is almost maximal
($\omega = \pi/4$), the conversions $\nu_e \leftrightarrow \nu_s$
can take place through vacuum oscillations.
Thus in this model
one has the combined effect of the vacuum and SMA resonance with the 
KK states as already discussed in Sec.~\ref{2-gen}.

The CHOOZ constraint is easily satisfied:  The $\nu_e$ survival probability 
contains an oscillatory term with the amplitude

\beq
\sin^2 2 \theta_e \approx 4 U_{e2}^2 U_{e3}^2 \eeq
which vanishes within our approximation. The LSND scale also gives an averaged 
contribution 

$$ 2 (U_{e4}^2+U_{e1}^2)(U_{e2}^2+U_{e3}^2)\sim \delta^4 $$
which is of ${\cal O}(10^{-4})$ and hence negligible.
 
The amplitude of the LSND probability 
$P(\bar\nu_\mu \to \bar\nu_e)$ is given by\footnote{
The contributions due to the KK states have been neglected
in  this approximation. Since from (\ref{B-alphan}), 
the coupling of $\nu_e$ and $\nu_\m$ to the KK states is
$\frac{\chi_e \chi_\mu}{n^2} \cdot \left(
\sum \frac{1}{k^2} \right)^{-1} \lsim
( U_{e2} U_{\m 2}+U_{e3}U_{\m3})$ and decreases
rapidly ($\propto 1/n^2$) with increasing $n$, this
approximation is valid up to a factor of ${\cal O}(1)$.}
\beq
P_{e\m}  \sim  4 (U_{e2} U_{\m 2}+U_{e3}U_{\m3})^2 
\sim  \chi_e^2 \chi_\mu^2~~. 
\label{lsnd-mumu}
\eeq
The LSND probability is significantly constrained here by
the solutions of the solar and atmospheric neutrino problems. 
The maximum allowed value of $\chi_\m^2\sim\chi_\tau^2$ for which
(\ref{dmsq-mumu}) reproduces the observed atmospheric scale 
is approximately given by $\chi_\m^2 \sim 0.02$. 
The $\chi_e^2$ is required to be around $3\cdot 10^{-4}$ 
to obtain the SMA solution to the solar neutrino problem. As a result,
$P_{e\m}\leq 10^{-5}$, which falls short of the LSND observation.

One can increase the LSND probability in the model by allowing
larger value for $\chi_e$ and thus by sacrificing the SMA solution.
This is seen from Fig.~\ref{lsn} 
which shows the effective LSND  mixing angle 
following from
(\ref{lsnd-mumu}) as a function of $\chi_e$. We have chosen
$\Delta m^2_{atm}$
and $\theta_{atm}$ at extreme values in the allowed range so as to 
maximize $\chi_\m$ and hence the value of the LSND probability in 
(\ref{lsnd-mumu}). 
It is seen from the figure that the
observed probability cannot be reproduced by the model for 
$\chi_e\leq 1$. The trend suggests that one may be able to 
obtain LSND probability if $\chi_e$ is chosen large. While the SMA solution
is no longer there, there is an alternative mechanism to solve the
solar neutrino problem here. 
Fig.~\ref{lsn} shows that 
$\dmsq_{14}$ increases with $\chi_e$ and eventually for 
$\chi_e\sim 0.5-0.9$ one obtains $\dmsq_{14}$ in the MSW range.
Thus instead of the higher excitations, the zero mode can cause
the MSW transition in this case.
The perturbative formalism followed here is no longer valid 
for $\chi_e\geq 1$ and it remains to be seen if all neutrino 
anomalies can be simultaneously understood  in this case.

\subsection{Three unmixed generations with Majorana masses
$\{ \m,\m,\m \}$}
\label{mumumu}

Starting with
the mass pattern $\mu,\mu,\mu$ for the active neutrinos, the
$4\times 4$ mass matrix involving $\nu_e,\nu_\m,\nu_\tau,\nu_s$ is given
by
\beq
m_L =\bmat{cccc}
\mu & 0 & 0 & m_e \\
0 & \mu & 0 & m_\mu \\
0 & 0 & \mu & m_\tau \\
m_e & m_\mu & m_\tau & 0 \emat~.
\label{ml-mumumo}
\eeq
Starting with this $m_L$ and including the corrections due
to seesaw approximation as in eq.(\ref{mlt-2}), 
the effective mass matrix is obtained as
\beq
\tilde{m}_L = \m\bmat{cccc}
1 - \chi_e^2 &  - \chie \chimu  & -\chie \chitau & \tau \chie \\
-\chie \chimu &  1 - \chimu^2   &  -\chimu \chitau &  \tau \chimu \\
-\chie \chitau & -\chimu \chitau &   1 - \chitau^2 &  \tau \chitau \\
\tau \chie &  \tau \chimu & \tau \chitau & 0  \emat~~.
\label{mlt-mumumu}
\eeq
The above matrix can be diagonalized exactly. The eigenvalues are given by
\beq
\lambda_1 = \mu ~,~
\lambda_2 = \mu ~,~
\lambda_{3,4} = {\mu\over 2}
\left(( 1 - \chi^2)\pm [(1-\chi^2)^2+4 \tau^2 \chi^2]^{1/2}
\right) ~,~
\label{lambda-mumumu}
\eeq
where $\chi^2=\chi_e^2+\chi_\m^2+\chi_\tau^2$.
This leads to the following (mass)$^2$ differences:
\beqa
\label{dmumumu}
\dmsq_{12}&=&0~,~ \nonumber \\
\dmsq_{13}&=&\dmsq_{23}\sim 2 \m^2 \chi^2~,~ \nonumber \\
\dmsq_{14}&=&\dmsq_{24}\sim \dmsq_{34}\approx \m^2 ~~.
\eeqa
The diagonalizing matrix is given up to terms quadratic in
$\{\chi_\a,\tau\}$ by
\beq
\hat{K} = R_{23}(\theta)~ R_{34}(\phi)~ R_{13}(\psi) ~R_{14}(\omega)~,
\label{k-mumumu}
\eeq
where the angles are given by 
\beq
\theta = \tan^{-1}\frac{\chimu}{\chitau} ~,~
\phi = - \tau \chiperp ~,~ 
\psi = \tan^{-1} \frac{\chie}{\chiperp} ~,~
\omega =  - \frac{\tau \chi_e \chi_\perp}{\chi^2} ~.
\label{angles-mumumu}
\eeq
This model contains two exactly degenerate states. As a result,
one has only two independent $\dmsq$s and it is not possible to account 
for all the anomalies. Moreover, the mixing pattern in (\ref{k-mumumu}) 
is such that
even if one is willing to give up LSND, the solar and atmospheric neutrino
anomalies cannot be simultaneously explained. In order to do this, 
one would need
to identify the larger $\dmsq \approx \m$ with the atmospheric neutrino scale.
But in that case, the atmospheric neutrino mixing angle following from
(\ref{k-mumumu}) turns out be too small 
[${\cal O}(s_\omega)$]. In spite of vanishing $\dmsq_{12}$, the solar
anomaly can be accounted  through resonance with the KK states. As
shown
in section (3), this becomes 
feasible  only if the electron neutrino
mass $\m\leq 10^{-3}\eV$. In this case, the 
$\Delta m^2$s in eq.(\ref{dmumumu}) cannot account for the atmospheric
neutrino anomaly.
Thus simultaneous explanation of the solar and atmospheric neutrino
is not possible and the model does not seem phenomenologically viable.

\section{Zee model for solving all anomalies}
\label{zee}

The neutrinos in the brane were assumed unmixed and degenerate
so far. Neutrino oscillations occurred entirely due to the presence
of coupling to the bulk neutrino. 
Since that seems to be inadequate for explaining all the 
neutrino anomalies, we now consider a more general
possibility in which some of the brane neutrino mixings
are present even in the absence of
the bulk states. The latter can provide additional structure needed to
understand all neutrino anomalies. This possibility will make some of the
simple neutrino mass generation mechanisms viable which by themselves
cannot
solve all the neutrino anomalies. This is exemplified by the model
due to Zee \cite{zee}. 
We confine our discussion to the Zee model although the basic
formalism in Sec.~\ref{formalism} can be used for any arbitrary 
mass structure in the brane.

One obtains \cite{st} the following  neutrino mass matrix in the Zee model:
\beq
m_0=\m\left( \ba{ccc}
0&\epsilon& s_\th \\
\epsilon&0&c_\th\\
s_\th& c_\th&0 \\ \ea \right) ~.
\label{m0-z} 
\eeq
where
\beq
\tan\th={f_{e\t}\over f_{\m\t}}~~~~~;~~~~~\e={f_{e\m}\over
f_{e\t}^2+f_{\m\tau}^2}\left({m_\m\over m_\t}\right)^2 ~~. 
\label{zeechoice} \eeq
The $f_{\a\b}$ in the above equation are Yukawa couplings of the charged
singlet Higgs to the leptonic doublet and $\m$ is the overall mass scale.

The above structure has been used to simultaneously solve
the solar and atmospheric \cite{zeesolar} or the LSND and the atmospheric 
neutrino anomalies \cite{st,bds}. The non-hierarchical $f$ imply
a very small $\e$ in eq.(\ref{m0-z}), 
leading to an approximate $L_e+ L_\m-L_\t$
symmetry. This corresponds to maximally mixed degenerate pairs.
Only $\nu_\m-\nu_e$ oscillations occur in this limit and the LSND result
can be explained  by choosing
$4 s_\th^2\approx 10^{-3}$ and $\m^2\sim 0.1-1 \eV^2$. Non-zero
$\e$ can generate spitting of ${\cal O}(4 \mu^2 \e s_\th)$ 
between the degenerate pairs.
This would correspond to the atmospheric
neutrino scale if $\e\sim 10^{-1}$. While small values for $\e$ are
more  natural in
this model, the required magnitudes of $\e,s_{\th}$ can be possible with
somewhat inverted hierarchy $f_{e\tau}\ll f_{\m\tau}\ll f_{e\m}$.
The solar neutrino problem cannot be accommodated in the Zee model with this
choice of parameters. 
This however becomes possible once coupling to  bulk neutrino 
is switched on. This coupling  also allows generation of the atmospheric 
neutrino scale even
when $\e$ is zero, i.e. model is $L_e+L_\m-L_\tau$ symmetric. 

The coupling of Zee model to the bulk neutrino leads to the
following mass matrix $m_L$ in the basis
$\nu_e,\nu_\m,\mu_\t,\nu_s$ to zeroth order in the seesaw approximation:
\beq
m_L = \mu \bmat{cccc}
0 & \eps & s_\th & r_e \\
\eps & 0 & c_\th & r_\mu \\
s_\th & c_\th & 0 & r_\tau \\
r_e & r_\mu & r_\tau & 0 \emat~.
\label{ml-z}
\eeq
Including non-leading corrections, we have the following 
effective mass matrix for the four neutrino states:
\barr \d m_L & & =  -{\m \over 2} \times \nonumber \\
 && \left( \ba{cccc}
2 (\chi_e\chi_\tau s_\th+\e\chi_e \chi_\m) &\chi_\tau X_{e\m}+\e Y_{e\m}&
s_\th Y_{e\t}+c_\th \chi_e\chi_\m+\e \chi_\m \chi_\t&
\chi_e  \b\\
\chi_\tau X_{e\m}+\e Y_{e\m}&
2(c_\th\chi_\m\chi_\t+\e \chi_e\chi_\m)& c_\th
Y_{\m\t}+s_\th\chi_e\chi_\m+\e \chi_e \chi_\tau&
\chi_\m  \b\\
s_\th Y_{e\t}+c_\th \chi_e\chi_\m +\e \chi_\m \chi_\t &
c_\th Y_{\m\t}+s_\th\chi_e\chi_\m+\e \chi_e \chi_\tau&
2 (\chi_\tau\chi_e s_\th+\chi_\tau\chi_\m c_\th)&
\chi_\t  \b\\
\chi_e  \b&\chi_\m  \b&\chi_\tau \b&0\\ \ea
\right) ,
\label{dml-lll} \earr 
where
\beqa
X_{\a\b}&=&c_\th\chi_\a+s_\th\chi_\b ~~,~~
Y_{\a\b}=\chi_\a^2+\chi_\b^2 ~~,
\nonumber \\
r_\a&=& m_\a / \m ~~,~~ \b= (\chi_e r_e+\chi_\m r_\m+\chi_\t
r_\t) ~.
\eeqa

Note that for the natural values $m_\a\sim 10^{-5}\eV$ and for $\m\sim
\eV$, the terms proportional to $\b$ are much smaller than other
elements in the matrix and can be neglected. Then up to the 
second
order in other parameters, $\tilde{m}_L$ can be written as:
\beq
\tilde{m}_L = \bmat{cccc}
0 & \eps -\chie \chitau/2 & s_\th - \chie \chimu/2 & 0 \\
\eps -\chie \chitau/2 & -\chimu \chitau &  1 - s_\th^2/2 - \chiperp^2/2 & 0\\
s_\th - \chie \chimu/2  &  1 -  s_\th^2/2 - \chiperp^2/2 
& - \chimu \chitau & 0 \\
0 & 0 & 0 & 0 \emat~.
\label{mlt-z}
\eeq 
This matrix can be diagonalized through
\beq
\hat{K} = R_{23}(\frac{\pi}{4})~
R_{13}(\frac{s_\th+\eps}{\sqrt{2}} - \frac{\chie \chipos}{2})~ 
R_{12}(\frac{s_\th-\eps}{\sqrt{2}} - \frac{\chie \chineg}{2}) ~
R_{23}(\frac{\eps^2-s_\th^2}{4})~,
\label{k-z}
\eeq
where $\chi_\pm \equiv (\chi_\mu \pm \chi_\tau)/\sqrt{2}$.
The rotation matrix ($\hat{K}$) may be expanded as
\beq
\bmat{cccc}
1 - \frac{\eps^2 + s_\th^2}{2} &  
\frac{ s_\th - \eps}{\sqrt{2}} - \frac{\chie \chineg}{2} &
\frac{\eps + s_\th}{\sqrt{2}} - \frac{\chie \chipos}{2} & 0 \\
- s_\th + \frac{\chie \chimu}{2} & 
\frac{1}{\sqrt{2}} (1 - \frac{s_\th^2}{2} + \frac{\eps s_\th}{2}) &
\frac{1}{\sqrt{2}} (1 - \frac{s_\th^2}{2} - \frac{\eps s_\th}{2}) & 0 \\
- \eps + \frac{\chie \chitau}{2} &          
\frac{1}{\sqrt{2}} (-1 + \frac{\eps^2}{2}  - \frac{\eps s_\th}{2}) &         
\frac{1}{\sqrt{2}} (-1 + \frac{\eps^2}{2}  - \frac{\eps s_\th}{2}) & 0 \\
0 & 0 & 0 & 1 \emat~.
\label{kfull-z}         
\eeq
The mass eigenvalues are 
\beq
\lambda_1 = -2 \mu \eps s_\th~ ,
\lambda_2 = \mu (-1 + \chineg^2 + \eps s_\th - \eps^2/2) ~,
\lambda_3 = \mu( 1 - \chipos^2 + \eps s_\th + \eps^2/2 ) ~,
\lambda_4 = 0.
\label{lambda-z}
\eeq
Thus, the three mass squared differences are
\beqa
\dmsq_{24} &\approx& \dmsq_{34} \approx \dmsq_{21} \approx
\dmsq_{31} \approx \mu^2 ~,~ \nonumber \\
\dmsq_{23} &\approx&  4 \mu^2 (\chimu \chitau + \eps s_\th) ~,~ \nonumber
\\
\dmsq_{14}&\approx& 4 \mu^2 \eps^2 s_\th^2~.
\label{dmsq-z}
\eeqa
The mass pattern is similar to the conventional 2+2 scheme.
This allows the solution to all anomalies
simultaneously:
The amplitude of the LSND oscillations is given by
\beq
P_{e\m} \sim 4 (U_{e2} U_{\mu 2}+ U_{e3} U_{\mu 3})^2 \approx 4 s_\th^2 ~~.
\label{lsnd-z}
\eeq 
Thus, the choice of $s_\th$ made at the beginning of the 
section is a suitable one.
All that is needed is $\mu^2 = \dmsq_{LSND}$.
The large mixing required for the atmospheric mixing
is naturally obtained: here
\beq
\sin^2 2\theta_{atm} \approx 4 U_{\mu_2}^2 U_{\mu_3}^2
\approx 1 - 2 s_\th^2~~,
\label{atm-z}
\eeq
so that the mixing is nearly maximal, and 
the mass squared difference is 
\beq
4 \mu^2 (\chimu \chitau + \epsilon s_\th) = \dmsq_{atm} \approx
3 \times 10^{-3} \eV^2~.
\label{atm2-z}
\eeq
Note that (\ref{lsnd-z}) and (\ref{atm-z}) do not constrain 
the value of $\e$ in any way, but (\ref{atm2-z}) restricts
the maximum value that $\e$ can take: if the major contribution
to $\dmsq_{atm}$ is from the $\e$ term, we have
$\e \sim \dmsq_{atm}/(4 \m^2 s_\th) \sim 10^{-2} - 10^{-1}.$ 
The $\nu_e$ survival probability contains an oscillatory term with the
amplitude
\beq
\sin^2 2\theta_e \approx 4 U_{e2}^2 U_{e3}^2 \approx 
(s_\th^2 - \e^2)^2  \lsim 10^{-2}~. 
\label{chooz-z}
\eeq
The LSND scale also adds an average term to this probability with the
amplitude $2 (\eps^2+s_\th^2)$.
Both these amplitudes are within the observed bounds in the CHOOZ experiment.
Notice that this experiment  puts a similar upper bound
on the allowed value of $\e$ as the atmospheric $\dmsq$.

For $\e\sim 0.1$ and $s_\th\sim 10^{-2}$, the  smallest mass difference
$\dmsq_{14}\sim 10^{-5}-10^{-6}\eV^2$. While this is the right value
for the MSW effect, the resonance cannot occur with the massless mode due
to the fact that the electron neutrino is heavier. 
The $\nu_e$ can however resonate
with the KK excitations with mass ${n\over R}>\lambda_1$. Thus 
the solar neutrino problem can be solved as in \cite{ds},
with quantitative details differing due to 
the presence of non-zero mass $\lambda_1$.
For $\e\sim 0.1$ and $\theta\sim 0.01$, this  mass
is in the MSW range and this can significantly alter the survival 
probability in the manner discussed in Sec.~\ref{nuemass}.

Let us now consider the  limiting case of the Zee model obtained when
$\e  \to 0$. As already mentioned, the mass matrix is invariant under
$L_e+ L_\m-L_\t$ in this case and implies a degenerate $\nu_e-\nu_\m$
pair. Such a matrix cannot lead to the atmospheric neutrino scale.
This scale is generated through the couplings to bulk neutrinos which
violate $L_e+ L_\m-L_\t$  symmetry. The expressions for the mixing
and masses can be recovered from the earlier case by putting $\e=0$.
This limit does not affect the solution to neutrino anomalies. One can
simultaneously solve all anomalies for the similar values of parameters
$R,s_\th,\m,\chi_a$ as in the case with non-zero $\e$. The electron 
neutrino mass now is much smaller than the MSW scale. 
As a result, the solution to solar neutrino 
problem occurs exactly as in \cite{ds} 
with very little perturbation from $\lambda_1$.

\section{Summary}
\label{concl}

We have analyzed the neutrino mass spectra in models
based on higher dimensional theories with an extra dimension of $mm$
size.  We have restricted our discussion to 
the minimal scenario where a single fermion
propagating in the bulk couples weakly to the
flavour neutrinos in the brane.
This coupling can generate and / or modify the masses and mixings
in flavour neutrinos. In addition,
the singlet fermion in the bulk provides a massless sterile
neutrino and a KK tower of several light sterile states which can
contribute to neutrino oscillations.
We have developed a formalism for calculating the net neutrino 
mass spectrum, taking into account the effect of the brane-bulk
coupling.
Simple approximations allow one to discuss various features of
the neutrino mass spectrum analytically.
We examined the possibility of solving all the neutrino anomalies ---
atmospheric, solar and LSND --- through this in a natural manner.

We have calculated the masses and mixings for several cases that can 
potentially solve all the neutrino anomalies.
In the absence of any masses for the brane neutrinos, it is not
possible to generate the required 
masses and large mixings naturally just
through the coupling with the bulk fermion. If two or more neutrinos
are massive and degenerate, however, large mixings and 
hierarchical mass splittings are possible. 
We considered two cases in detail, the one in which all three
flavour neutrinos are degenerate, and the one in which two
of them are degenerate and the third massless.
It turns out that if all the three neutrinos are degenerate,
their coupling to the bulk fermion cannot lead to 
a simultaneous understanding of even the solar and atmospheric 
neutrino anomalies, let alone LSND in addition. This leaves the
other option as the only viable alternative.

When two of the three neutrinos are degenerate with mass $\m$ 
and the neutrinos have no a priori mixings among themselves, 
their couplings to the bulk fermion automatically 
lead to two pairs of almost degenerate neutrinos separated by the scale
characterized by $\m$. The splittings within the pair are hierarchical,
and may account for $\dmsq_\odot$ and $\dmsq_{atm}$.
The hierarchy is completely
controlled by the higher dimensional physics and one typically finds 
$\dmsq_\odot / \dmsq_{atm} \sim \delta^4$ where 
$\delta$ refers to the 
typical mixing between the bulk and brane neutrinos. 
This hierarchy generates the scale corresponding to 
long wavelength solar oscillations for $\delta^2 \sim 10^{-2}$.  
Identification of $\m$ with the LSND scale 
turns out to generate the atmospheric scale for the same value of $\delta$.
Thus the features needed to account for the solar and atmospheric 
neutrino anomalies follow automatically. 
But the $\nu_e \leftrightarrow \nu_\m$ oscillation probability is
much smaller than that observed at LSND  when the mixing $\delta$ is small.

Understanding of all neutrino anomalies would then require 
some preexisting mass structure among the flavour neutrinos. 
We analyzed the specific example of the neutrino mass matrix
predicted by the Zee model, which produces two massive degenerate
neutrinos and a massless one.
The coupling with the bulk fermion leads to the 2+2 mass spectrum,
and its solutions to solar and atmospheric neutrino 
anomalies can coexist with large $\nu_e-\nu_\m$ oscillation probability 
seen at LSND.
Even in the $\epsilon \to 0$ limit of the Zee model
(which corresponds to the $L_e + L_\mu - L_\tau$ symmetry), all the
three neutrino anomalies can be naturally accounted for.
The LSND scale $\mu$, the LSND mixing angle $\theta$ and the large 
atmospheric mixing angle are already present in the structure
of the Zee (or $L_e + L_\mu - L_\tau$ symmetric) mass matrix.
The bulk fermion provides a massless sterile neutrino and a KK tower
of sterile states that can participate in the neutrino oscillations.
The size of extra dimensions $R \sim mm$ creates masses of the lighter 
KK states in the range $m_n^2 \sim 10^{-6}$ eV$^2$, appropriate
for the SMA MSW solution of the solar neutrinos.
The brane-bulk coupling also generates the splitting
$\dmsq_{atm} \sim 4 \mu^2 \xi^2$ and the solar mixing angle
$\sin \theta_\odot \sim \xi_e/n$. The addition of the bulk
fermion and its coupling simultaneously provides
$\dmsq_{atm}, \dmsq_\odot$ and $\theta_\odot$ in the right range,
thus completing the picture. The Zee model embedded in
extra dimensions in the minimal way can then account for all the
neutrino anomalies.

The sterile neutrinos participate in the solar neutrino oscillations.
We have calculated how the survival probabilities in \cite{ds}
get modified due to the nonzero masses of $\nu_e$ that is obtained
in the preferred scenarios.
Another interesting feature of some of the scenarios considered here is the 
simultaneous occurrence of two different mechanisms for the solution to 
the solar neutrino problem. The 
oscillations of $\nu_e$ to massless mode of the bulk state corresponds to 
the long wavelength solution and oscillations to higher modes are 
appropriate for the SMA MSW conversion. These lead to differences 
compared to the phenomenology of the conventional 
``2+2'' schemes \cite{22} 
and allows one to explain \cite{msw+vac} various features of the solar 
neutrino data.

We had assumed throughout that size of the extra dimension and the 
fundamental scale are near their observable limits. It is seen from the
present considerations that this observability does not conflict with
the observed features of neutrino masses and mixings. While the presence of 
large extra dimensions cannot exclusively account for all the neutrino
anomalies they can provide an important ingredient for generating the observed 
features of the neutrino spectrum.

\section*{Acknowledgments}
We would like to thank the Theory Division at CERN for 
hospitality where part of this work was done.

\end{document}